\documentclass[reprint, amsmath,amssymb,aps,pra]{revtex4-2}

\usepackage{graphicx}
\usepackage{dcolumn}
\usepackage{bm}
\usepackage{amsmath}
\usepackage{svg}
\usepackage{url}
\usepackage{hyperref}
\usepackage[mathlines]{lineno}
\usepackage{xcolor}
\usepackage{placeins}
\usepackage{multirow}
\usepackage{comment}

\begin{document}

\title{Quantum Time Synchronization of Star Networks}
\author{Brian J. Rollick}
\email{b.rollick@cablelabs.com}
\affiliation{CableLabs, 858 Coal Creek Circle, Louisville, CO, 80027}

\author{Bernardo A. Huberman}
\email{bahuberman@gmail.com}
\affiliation{RISE - Research Institutes of Sweden, Stockholm, Sweden}

\author{Zhensheng Jia}
\email{s.jia@cablelabs.com}
\affiliation{CableLabs, 858 Coal Creek Circle, Louisville, CO, 80027}

\date{\today}

\begin{abstract}
We extend the single-source approach of Valencia et al. in order to synchronize the clocks of an N-user star network, connected both through fiber and in free space. Entangled photon pairs from a centralized SPDC source are distributed through a 1xN splitter to four remote users arranged in a star topology. Using commercially available single-photon detectors and time-taggers, we achieve median timing precision of 50 ps for atomic oscillators and 20 ps for GPS disciplined oscillators in our Kalman models. Thus, we achieve a three order of magnitude improvement over GPS alone. By monitoring the drift of the correlation peaks over time, we also extract the frequency skew between users' local clocks to 35 ps/s precision. From these measurements, each user can compute its offset and drift relative to every other user, achieving full network synchronization without a central clock.
\end{abstract}

\maketitle

\section{Introduction}

Reliable and precise time synchronization is a keystone technology of modern society. Without it, global communications, financial systems, and scientific infrastructure would not function coherently. It underpins applications as diverse as navigation systems, fundamental tests of relativity, and distributed fiber sensing.

Several methods provide scalable time synchronization for today’s technological landscape. For internet-connected systems, the Network Time Protocol (NTP) \cite{Mills1991} is the most common procedure used, while users requiring higher precision often employ the Precision Time Protocol (PTP) (IEEE1588-2019).

NTP typically achieves millisecond-level precision, and is limited by variable network latency, whereas PTP can reach the sub-microsecond regime under ideal conditions. The DOCSIS Time Protocol (DTP) \cite{AndreoliFang2018} mitigates latency by allocating a dedicated synchronization channel.
In addition, because the Earth’s surface is bathed in GPS signals \cite{Lewandowski1999}, GPS remains a convenient timing reference for devices lacking internet connectivity.

Specialized domains such as high-frequency trading and large-scale scientific experiments (e.g., CERN) demand tighter precision. They employ a hardware-specific extension of PTP known as the White Rabbit Time Protocol (WRTP) \cite{Lipinski2011}, achieving clock alignment in a few nanoseconds.

Despite these advances, classical protocols remain vulnerable to spoofing and delay attacks, as timing information can be copied or manipulated in transit. Indeed, successive versions of PTP have required security patches as new attacks are discovered \cite{Han2019, Alghamdi2022, Itkin2020, Ullmann2009}.
Motivated by the non-cloning nature of quantum states and their inherent resistance to eavesdropping, numerous groups have proposed quantum time synchronization (QTS) schemes \cite{Chuang2000, Jozsa2000, Giovannetti2001, Giovannetti2004, Kong2017, Ren2012, Nande2025}.

Most QTS protocols exploit the near-simultaneous generation of entangled photon pairs in spontaneous parametric down-conversion (SPDC) \cite{Burnham1970}. In this nonlinear $\chi^{(2)}$ process, a pump photon is converted into two lower-energy photons \cite{Boyd2020}.
Users can then compute the second-order correlation function \cite{Glauber1963} between remote detectors to identify photons belonging to the same pair. However, both the required histogram and the cross-correlation function can be computationally intensive, imposing constraints on local clock stability. Several works \cite{Ho2009, Spiess2023a, Spiess2023, Quan2016} have proposed improved post-processing to relax these requirements.

Other protocols use two SPDC sources \cite{Lee2019, Crum2025, Hong2021, Hong2024, Hou2019} to cancel channel-dependent optical delays, making them attractive for satellite applications. Such dual-source architectures significantly reduce asymmetry-based timing errors, although asymmetric-delay attacks remain possible \cite{Lee2019a}.
A related approach \cite{DAuria2020} drives multiple remote SPDC sources with a common classical pump and employs Hong–Ou–Mandel (HOM) interference for clock alignment.

In contrast, single-source schemes offer architectural simplicity, as users need not operate their own photon sources. Valencia et al. \cite{Valencia2004} demonstrated this by distributing entangled pairs between two nodes. Although such systems are sensitive to path-length uncertainty, they can nevertheless reach femtosecond-level precision \cite{Quan2019}.

Many quantum-synchronization demonstrations rely on satellite links \cite{Ducoing2025, Komar2014, Dai2020}, limiting their use in indoor or underground environments. Moreover, scalability to many users is crucial for realistic networks.

In this work, we extend the single-source approach of Valencia et al. to an N-user star fiber network, as well as a free space one. We distributed entangled photon pairs from a centralized SPDC source through a 1×N splitter to four remote users arranged in a star topology. Using commercially available single-photon avalanche detectors and time-taggers, we achieved timing precision between atomic oscillators of about 50 ps over 30 seconds, and 20 ps over 0 seconds when piggybacked on GPS disciplined oscillators (GPSDO). We extracted these offset precisions using a Kalman model of repeated offset measurements and assuming constant skew.

In addition, by monitoring the drift of the correlation peaks over time, we extracted the frequency skew between users’ local clocks. From these measurements, each user can then compute its offset and drift relative to every other user, achieving full network synchronization without a central clock.

This paper is organized as follows: Section II reviews the physical principles underlying our techniques and approach. Section III details the experimental setup and data-processing workflow. Section IV presents and discusses the results. We discuss our Kalman model used to extract precision in the Appendix.

We should mention that while all our results were mostly obtained in a fiber star network, our free space measurements indicate that they can easily be extended to signal propagation in free space. In principle, this should yield higher precision as the dispersion of optical fibers is avoided. This would render this method useful for synchronizing LEO satellite constellations\cite{Sandholm2026}.

\section{BACKGROUND}

Spontaneous Parametric Downconversion (SPDC) is a standard technique in quantum optics and quantum information experiments \cite{Kwiat1995}. In SPDC, a strong laser pumps a second-order ($\chi^{(2)}$) nonlinear crystal that is periodically poled to satisfy phase-matching requirements between the crystal’s optical axes.

In the undepleted-pump approximation, the pump field is treated as a classical coherent drive, while the signal and idler fields are quantized. The resulting interaction Hamiltonian is:

\begin{equation}
\label{SPDC_process}
\hat{H}_{\mathrm{SPDC}}
= 
\varepsilon_{0}
\int_{V} d^{3}\mathbf{r}\;
\chi^{(2)}(\mathbf{r})
\,\mathcal{E}_{p}^{(+)}
\,\hat{E}_{s}^{(-)}
\,\hat{E}_{i}^{(-)}
+\mathrm{h.c.}
\end{equation}

Here, a pump photon, indicated by the $\mathcal{E}_{p}^{(+)}$, can spontaneously generate a pair of lower-energy photons, shown by $\hat{E}_{s,i}^{(-)}$, of approximately twice the wavelength. The conversion probability is low, so under continuous-wave (CW) pumping the photon pairs are sparsely distributed in the time domain. Current state-of-the-art SPDC sources produce on the order of $10^{6}$ pairs per second; if these pairs are well spread out in time, one would expect roughly $1~\mu\mathrm{s}$ of separation between successive pairs on average.

Our technique relies only on the fact that SPDC produces two photons at a time. Although the generated pairs are polarization-entangled, we do not use polarization information; SPDC is simply the most practical and efficient mechanism for producing correlated photon pairs in our setting. Alternatives such as weak coherent pulses (used in decoy-state QKD) frequently contain pulses with zero or one photon, while many single-photon sources such as quantum dots still suffer from low extraction efficiency, tending to yield one-photon pulses. Thus, SPDC remains the best available method for generating simultaneous photon pairs in our work.

\begin{figure}
    \centering
    \includegraphics[width=0.5\linewidth]{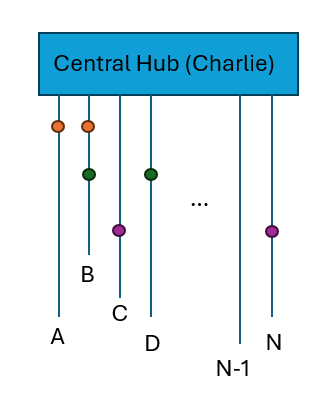}
    \caption{A central hub (Charlie) produces entangled photons and the photons are split up using an N splitter. Most likely, the photons leave separate ports of the splitter and two photons reach different users. In the figure, the pieces of the same photon pair are indicated by the same color. It should be noted, that the photon pairs are much further spaced in the time domain.}
    \label{fig:splittingDiagram}
\end{figure}

A central hub generates photon pairs, which then propagate through a feeder fiber of up to $10~\mathrm{km}$. Upon reaching a $N$-port splitter, the pair is randomly routed among the output ports. With probability $1 - 1/N$, the two photons exit through different ports and therefore reach different users in a $N$-user star network. A visualization is shown in Figure ~\ref{fig:splittingDiagram}. While many works treat the splitter as loss, here it serves purely as a distribution mechanism for routing photon pairs among users. If we intended to synchronize only two users, the effective splitter loss would matter; however, since we wish to synchronize all users, any splitting event in which the two photons reach different users would be productive. Only when both photons reach the same user does an event become uninformative.

Each user operates one (or more, if polarization measurements are used) single-photon detector. The users’ time-tagging electronics may have different resolutions and timing jitter, and all users are assumed to operate asynchronously. After sufficient data has been collected, each user sends their detection timestamps to the central hub, which then broadcasts them to all users.

Because photons from the same SPDC pair arrive within a narrow, correlated time window, the users can estimate their relative clock offsets by measuring the second-order correlation function \cite{Glauber1963}:

\begin{equation}
\label{Glauber}
g^{(2)}(t_1,t_2)
 = \frac{
     \left\langle
       \hat{E}^{(-)}(t_1)\,
       \hat{E}^{(-)}(t_2)\,
       \hat{E}^{(+)}(t_2)\,
       \hat{E}^{(+)}(t_1)
     \right\rangle
   }{
     \left\langle
       \hat{E}^{(-)}(t_1)\hat{E}^{(+)}(t_1)
     \right\rangle
     \left\langle
       \hat{E}^{(-)}(t_2)\hat{E}^{(+)}(t_2)
     \right\rangle
   }.
\end{equation}

Experimentally, we approximate the intensity operators by point processes of detection events,

\begin{equation}
I_A(t) = \sum_i \delta\!\bigl(t - t_i^{(A)}\bigr), \qquad
I_B(t) = \sum_j \delta\!\bigl(t - t_j^{(B)}\bigr),
\end{equation}

and evaluate the correlation by histogramming the time differences $\Delta t = t_j^{(B)} - t_i^{(A)}$. We restrict attention to coincidences satisfying $|\Delta t| \le \tau_{\mathrm{win}}$, where $\tau_{\mathrm{win}}$ is chosen based on the dominant timing uncertainties (detector jitter, electronic noise, and fiber dispersion). Using histogram bins of width $\Delta\tau$, the discrete estimator of the normalized correlation is

\begin{equation}
g^{(2)}_{AB}(\tau_k)
\simeq
\frac{C_k}{R_A R_B\, T\, \Delta\tau},
\end{equation}

where $C_k$ is the number of coincidence events in the $k$-th bin, $R_A = N_A/T$ and $R_B = N_B/T$ are the singles rates, and $T$ is the total acquisition time. The clock offset between channels $A$ and $B$ is obtained from the delay $\tau_k$ at which $g^{(2)}_{AB}(\tau_k)$ (equivalently $C_k$) attains its maximum.

In practice, the offset we calculate combines both the fiber delay and the true clock offset of the user. However, since we are working in a star fiber network, we assume the positions are fixed. Indeed, the distances can be known very accurately using classical methods. In the general case, the calculated $\tau_k = t_{AB} + \delta_{AB}$ where t is the propagation time of the signal and $\delta_{AB}$ is the true clock offset.

Given a short, noisy measurement, it is possible that the true offset may not yield the maximum g2. We introduce a method for filtering out these false peaks called Triangular Closure. Here, we take advantage of the fact that:

\begin{equation}
    \Delta_{ABC} = \delta_{AC} + \delta_{CB} - \delta_{AB}
\end{equation}

Where the $\delta$ represents the individual offsets between clocks. The total offset between three pairs of three clocks, $\Delta$ will be close to zero. When a false value for tau gives a maximum g2 measurement, the $\Delta$ will exceed some statistical threshold and the false peak can be ignored or downweighted heavily in its Kalman model. 

Upon finding tau which maximized g2, we track the movement of the peak using a Kalman model, where the state consists of the offset and the frequency skew between measurements. Kalman modeling was chosen because it provided a reasonably robust way to compute offset and skew when measurements have different amounts of noise. The model provides varying statistical weights to measurement times when uncertainty is lowest. Our model is described in detail in the Appendix. 

\section{EXPERIMENT}

\begin{figure*}
    \centering
    \includegraphics[width=1.0\linewidth]{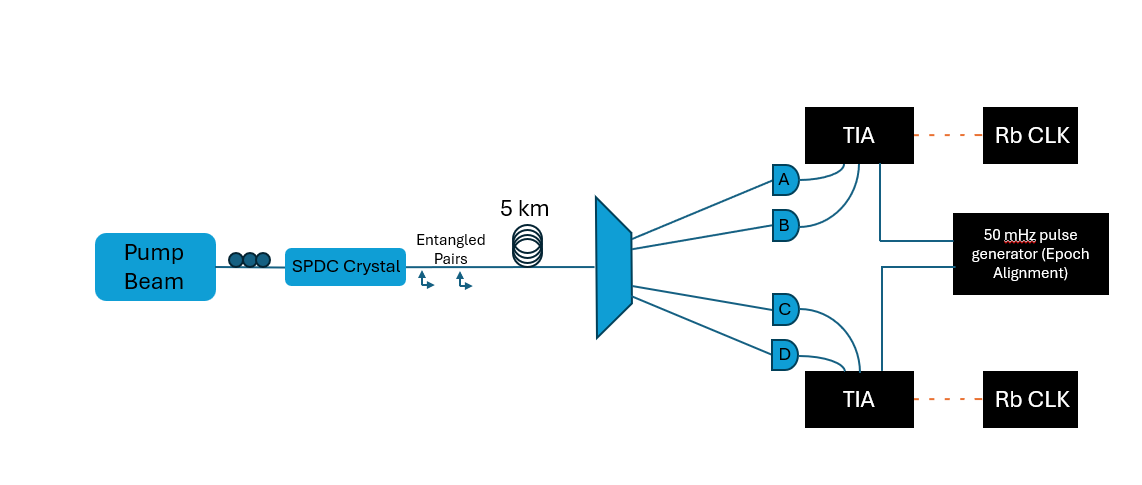}
    \caption{A central hub produces entangled photon pairs and sends them to an N=8 splitter. In this case, four ends are terminated. The other four go to any of four detectors located between two stations. The two stations have separate clocks, but the detectors on a single station share the same time tagger and therefore clock. We use a single pulse for epoch alignment due to the difficulty in beginning the two separate TIA's recordings at the same time. In the field where this would happen continuously, epoch alignment is likely unnecessary.}
    \label{fig:apparatus}
\end{figure*}

A diagram of our experimental setup can be seen in Figure ~\ref{fig:apparatus}.

In our experiment, we chose to use an off-the-shelf nonlinear crystal (Optilab) with temperature control in order to ensure that degenerate photons at 1560 nm are produced. At optimal temperature, the photons had a bandwidth of about 1 nm FWHM. 

We pump the crystal with a narrow band laser (IPS) centered at 780.2 nm. The pump laser is temperature controlled to minimize polarization drift. We filtered the output to remove harmonic noise from the laser, which could produce noise in the same band as the entangled photons. The polarization was adjusted prior to reaching the crystal so that the count rates are maximized for type II SPDC. In this work, we produce degenerate photon pairs in order to minimize dispersion in a potentially long feeder fiber. The nonlinear crystal is temperature-stabilized at $61.15^\circ\mathrm{C}$ to achieve the birefringence needed for phase matching.

We cascaded two filters after the nonlinear crystal to remove the pump beam. Afterward, the photons are sent along a 5 km spool of feeder fiber. The photons then reach a 1 by 8 splitter, where four paths lead to free running single photon detectors (MPD and IDQ). The MPD detectors' detections are time tagged by a quTools time interval analyzer (TIA), while the IDQ detectors are tagged by an IDQ TIA. The TIA's have independent clocks. Some datasets are taken with off the shelf Rubidium atomic clocks with filtered output, and the others are GPS Disciplined oven controlled crystal Oscillators (GPSDO). 

Due to the difficulty in starting the TIA's recording at the same time, we used a signal to align the epochs of each. It is expected that in the field, this procedure would not be necessary because the synchronization can run asynchronously, so the epochs can be easily known from prior synchronizations. 

We can choose the lengths of datasets depending on the tradeoff between getting more single photon events, and thereby tighter uncertainty and more confidence in identifying the offset, with the potential smearing of the peak due to frequency differences between the clocks. 


For large numbers of photon pairs, the intensity distribution approximates the classical case and the users, depending on their detector efficiencies, will receive similar numbers of photons. For example, in a 32 user network, the splitter contributes 15 dB of loss. Every time the splitter size doubles, the number of photons received must remain the same to ensure the same accuracy. To achieve this end, the pump intensity at the central hub can either be increased or the users will need to record data for double the time. 

In the case of larger split ratios, it is expected that the users will employ Rb based atomic clocks or GPS steering rather than OXCO clocks to minimize frequency drift when datasets are longer.

We present our results in Tables I and II. The standard errors (SE) are those after propagating many single shot Gaussian measurements with FWHM approximately 100-200 ps. 

\begin{table}[ht]
\centering
\label{tab:atomic}
\begin{tabular}{lrrrr}
\hline
Pair & Offset (ps) & Offset SE (ps) & Skew (ps/s) & Skew SE (ps/s) \\
\hline
A-B & 25191        & 30.815 & 9.6866   & 33.155 \\
A-C & -256040  & 42.983 & -487.11  & 35.464 \\
A-D & -304820  & 104.06 & -536.19  & 45.76  \\
B-C & -281260  & 43.441 & -528.03  & 35.266 \\
B-D & -329670  & 49.456 & -472.52  & 34.943 \\
C-D & -48675       & 62.444 & -29.169  & 40.699 \\
\hline
\end{tabular}
\caption{Pairwise clock offset and skew estimates for Rb atomic oscillators}
\end{table}

\begin{table}[ht]
\centering
\label{tab:GPS}
\begin{tabular}{lrrrr}
\hline
Pair & Offset (ps) & Offset SE (ps) & Skew (ps/s) & Skew SE (ps/s) \\
\hline
A-B & 25315  & 13.515 & 104.23   & 57.303 \\
A-C & -40169 & 50.066 & -131.91  & 61.916 \\
A-D & 11282  & 14.54  & -175.54  & 62.189 \\
B-C & -65308 & 19.45  & -151.53  & 59.575 \\
B-D & -13916 & 19.967 & -90.234  & 57.934 \\
C-D & 51384  & 11.376 & -28.731  & 56.967 \\
\hline
\end{tabular}
\caption{Pairwise clock offset and skew estimates for GPS disciplined clocks.}
\end{table}

\section{DISCUSSION}

Our results from Kalman filtering are contained in Tables I and II for synchronization precision using atomic and GPSDO's, respectively. The offsets in the tables are the offsets of each clock pair at a given measurement time, and the skew is the speed at which the clocks drift apart relative to one another. In our experiment, there are only two independent clocks, so each of the skews is a measurement of the same value. 

We can estimate the quality of the model by comparing the residual of the triangle closure of the g2 maxima for a given triple, and the Kalman's predicted offsets for the said triple. The trends reveal the fact that skews require longer measurements and a reasonably large allowance for noise, but precise offset measurements get worse for very long Kalman predictions. This is due to the fact that we model offset as the integral of skew. Any noise not contained in the model for skew therefore grows as the model accumulates offset measurements. 

Uncertainty estimates reported for each link correspond to the Kalman posterior $1\sigma$ uncertainties obtained under a two-state (offset and fractional frequency) stochastic clock model, with measurement noise derived from Gaussian fits to the $g^{(2)}(\tau)$ correlation peaks. 

As an independent consistency check, we evaluated triangle-closure residuals formed from three simultaneously estimated links. For example, for the atomic case, we observe that the closure residual root-mean-square (RMS) values are typically larger than the median per-link Kalman uncertainties (e.g., $\sim 200\,\mathrm{ps}$ RMS closure versus $\sim 50\,\mathrm{ps}$ median per-link $1\sigma$). This behavior is expected because triangle closure combines three links and is therefore sensitive to inter-link correlations as well as occasional non-Gaussian measurement outliers (e.g., peak misidentification events).

Accordingly, we interpret the per-link Kalman uncertainties as model-conditioned internal precision, while triangle-closure residuals provide a more conservative system-level self-consistency metric.

In the GPSDO case, we observe a larger disagreement between the Kalman offset uncertainties and what we would expect from the Triangle Closure residual because the clock model assumes a constant skew. A more sophisticated, future model might include steering as a perturbation to the skew. 

Depending on the type of oscillator used for each user, it is beneficial to tune the parameters of the Kalman model, such as whether to use a "rolling window" where only more recent measurements are included in the offset model and the whole dataset is included in the skew. If users are primarily using GPSDO's, where slow steering occurs, the offsets are more stable over long measurements because skew noise is slowly compensated by the satellite reference (perhaps the same satellite reference for local clocks). Atomic oscillators, albeit extremely stable, contain skew noise that is not modeled in our work. Therefore, we use a rolling window of 30 seconds for datasets using atomic clocks, and no rolling window for GPSDO's. 

\subsection{SECURITY}




It is worthwhile to heuristically consider how an adversary might attempt to attack our protocol. Suppose that an attacker is outside the time-distribution network and tries to inject fake photons. Such an adversary has no way of knowing when true photon pairs will arrive or which users will receive them. Even if the attacker operates their own SPDC source and attempts to shift the signal or idler photon in time, it cannot control which users receive which photons because the splitter is entirely passive. Any injected light would therefore fail to reproduce the correlated detection statistics required to create a coincidence peak.

A more capable adversary might be inside the network and attempt to manipulate the protocol by reporting falsified detection times to the central hub. In this case however, every honest user still receives timing information from all other users. The redundant set of pairwise correlations also used in Triangular Closure exposes inconsistencies immediately, so a malicious participant cannot alter their timestamps in a way that produces a self-consistent set of clock offsets. The attacker gains nothing beyond revealing its own misbehavior.

The strongest adversary is one who controls the central hub itself. Such an adversary can distribute photon pairs and intercept or modify timing information before forwarding it to the users. But even here, the adversary faces fundamental limits. The users require genuine photon-pair correlations to estimate their offsets; without sending real pairs, no correlation peak will form, and any synchronization attempts will simply fail. If the malicious hub instead tries to modify the users’ timestamps, the inconsistency becomes visible because the passive splitter prevents the hub from selectively hiding or redirecting photons. While an adversary can jam the protocol by withholding or corrupting information, it cannot falsely convince the users that they are synchronized.

\section{CONCLUSION}

In this work, we reported a successful extension of the single SPDC source time synchronization method to many user star network. As we showed, we were able to synchronize clocks with a median time precision of 50 ps for atomic oscillators and 20 ps for GPS disciplined oscillators, using a Kalman model. By monitoring the drift of the correlation peaks over time, we also extracted the frequency skew between users' local clocks to 30-60 ps/s precision. From these measurements, each user can compute its offset and drift relative to every other user, achieving full network synchronization without a central clock. This is to be compared with the precision with which GPS systems presently operate, which is on the order of 10's of nanoseconds.

It is worth pointing out that these results were obtained using off the shelf equipment from different brands. A more sophisticated instrumentation, such as superconducting nanowire single photon detectors (SNSPD's) would lead to better synchronization and precision in determining the skew values. In addition, dispersion in the feeder fiber causes broadening of the photon pairs in time. Use of hollow core fibers, free space channels or narrow band SPDC sources could offer further improvement.

Although we focused on fiber communication, it is important to state that this methodology can also be applied to satellite based mesh systems. In addition, the improvements one obtains from working in free space translate into much better values for the synchronization and skews of the detectors. In this space scenario, individual satellites in the constellation can act as the SPDC source with neighboring satellites playing the role of user nodes. This in turn would ensure clock synchronization in a mesh constellation far more precise than what is achievable today.

Finally, we point out that we did not take advantage of any polarization correlations in this work. We merely used the SPDC as a convenient source of photon pairs tightly correlated in time. Future work might exploit polarization entanglement as an additional protection against spoofing, or even modify the detector setups for key exchange in addition to time synchronization. 

\begin{acknowledgments}
    The authors thank Jing Wang of CableLabs for helpful conversations. 
\end{acknowledgments}

\FloatBarrier

\appendix
\section{Kalman Filtering Model for Time Offset and Frequency Skew}
\label{app:kalman}

In this appendix, we describe the state-space model and estimation procedure used to track the relative time offset and fractional frequency skew between pairs of clocks using correlation-derived delay measurements.

\subsection{Measurement Model}

For a given clock pair, each correlation histogram yields an estimate of the relative time offset.
Let
\begin{equation}
z_k = x(t_k) + v_k
\end{equation}
denote the measured time offset at discrete time $t_k$, where:
\begin{itemize}
  \item $x(t)$ is the true time offset between the two clocks,
  \item $v_k$ is zero-mean measurement noise with variance $R_k$.
\end{itemize}

The measurement variance $R_k$ is obtained from the statistical uncertainty of the fitted correlation peak and is treated as known for each $k$.

In matrix form,
\begin{equation}
z_k = \mathbf{H}\mathbf{x}_k + v_k,
\qquad
\mathbf{H} = \begin{bmatrix} 1 & 0 \end{bmatrix}.
\end{equation}

\subsection{State Vector}

The system state is defined as
\begin{equation}
\mathbf{x}_k =
\begin{bmatrix}
x_k \\
y_k
\end{bmatrix},
\end{equation}
where:
\begin{itemize}
  \item $x_k$ is the time offset (s),
  \item $y_k$ is the fractional frequency offset (dimensionless), corresponding to the time derivative of the offset.
\end{itemize}

\subsection{Dynamical Model}

Between measurements, the time offset evolves according to
\begin{equation}
\frac{dx}{dt} = y(t),
\qquad
\frac{dy}{dt} = w_y(t),
\end{equation}
where $w_y(t)$ is a zero-mean white-noise process that models frequency wander.

The discretization of the interval $\Delta t_k = t_k - t_{k-1}$ yields the state transition model
\begin{equation}
\mathbf{x}_k = \mathbf{F}_k \mathbf{x}_{k-1} + \mathbf{w}_k,
\end{equation}
with
\begin{equation}
\mathbf{F}_k =
\begin{bmatrix}
1 & \Delta t_k \\
0 & 1
\end{bmatrix}.
\end{equation}

The process noise covariance is given by
\begin{equation}
\mathbf{Q}_k =
\begin{bmatrix}
\sigma_x^2 \Delta t_k & 0 \\
0 & \sigma_y^2 \Delta t_k
\end{bmatrix},
\end{equation}
where $\sigma_x$ and $\sigma_y$ characterize random-walk noise in offset and fractional frequency, respectively.

\subsection{Kalman Filter Equations}

At each time step, the filter executes a prediction step
\begin{align}
\hat{\mathbf{x}}^-_k &= \mathbf{F}_k \hat{\mathbf{x}}_{k-1}, \\
\mathbf{P}^-_k &= \mathbf{F}_k \mathbf{P}_{k-1} \mathbf{F}_k^\mathsf{T} + \mathbf{Q}_k,
\end{align}
followed, when a valid measurement is available, by an update step
\begin{align}
\mathbf{K}_k &= \mathbf{P}^-_k \mathbf{H}^\mathsf{T}
              \left(\mathbf{H}\mathbf{P}^-_k\mathbf{H}^\mathsf{T} + R_k\right)^{-1}, \\
\hat{\mathbf{x}}_k &= \hat{\mathbf{x}}^-_k + \mathbf{K}_k
                     \left(z_k - \mathbf{H}\hat{\mathbf{x}}^-_k\right), \\
\mathbf{P}_k &= (\mathbf{I} - \mathbf{K}_k \mathbf{H}) \mathbf{P}^-_k.
\end{align}

If no valid measurement is present at time $k$, only the prediction step is applied.

\subsection{Initialization and Sliding-Window Estimation}

To mitigate long gaps and nonstationary behavior, the filter is applied within a sliding temporal window of fixed duration.
For each window, the initial offset is set by the first valid measurement, and the initial fractional frequency offset is estimated from the slope between the first two valid measurements within the window.

Only the final state estimate at the window endpoint is retained, producing a time series of locally optimal offset and skew estimates.

\subsection{Uncertainty Estimates}

The reported $1\sigma$ uncertainties in offset and fractional frequency are obtained from the diagonal elements of the state covariance matrix,
\begin{equation}
\sigma_x(k) = \sqrt{P_{k,11}},
\qquad
\sigma_y(k) = \sqrt{P_{k,22}}.
\end{equation}

These uncertainties propagate naturally into network-level consistency checks, such as triangle closure residuals, described elsewhere in the text.

\bibliography{TimeSynchronization}

@Article{Valencia2004,
  author   = {Valencia, Alejandra and Scarcelli, Giuliano and Shih, Yanhua},
  journal  = {Applied Physics Letters},
  title    = {Distant {Clock} {Synchronization} {Using} {Entangled} {Photon} {Pairs}},
  year     = {2004},
  issn     = {0003-6951, 1077-3118},
  month    = sep,
  note     = {arXiv:quant-ph/0407204},
  number   = {13},
  pages    = {2655--2657},
  volume   = {85},
  doi      = {10.1063/1.1797561},
  url      = {http://arxiv.org/abs/quant-ph/0407204},
  urldate  = {2025-03-07},
}

@Article{Lee2019,
  author   = {Lee, Jianwei and Shen, Lijiong and Cerè, Alessandro and Troupe, James and Lamas-Linares, Antia and Kurtsiefer, Christian},
  journal  = {Applied Physics Letters},
  title    = {Symmetrical clock synchronization with time-correlated photon pairs},
  year     = {2019},
  issn     = {0003-6951, 1077-3118},
  month    = mar,
  note     = {arXiv:1812.08450 [quant-ph]},
  number   = {10},
  pages    = {101102},
  volume   = {114},
  doi      = {10.1063/1.5086493},
  url      = {http://arxiv.org/abs/1812.08450},
  urldate  = {2025-03-07},
}

@Article{Hou2019,
  author    = {Hou, Feiyan and Quan, Runai and Dong, Ruifang and Xiang, Xiao and Li, Baihong and Liu, Tao and Yang, Xiaoyan and Li, Hao and You, Lixing and Wang, Zhen and Zhang, Shougang},
  journal   = {Physical Review A},
  title     = {Fiber-optic two-way quantum time transfer with frequency-entangled pulses},
  year      = {2019},
  issn      = {2469-9934},
  month     = aug,
  number    = {2},
  pages     = {023849},
  volume    = {100},
  doi       = {10.1103/physreva.100.023849},
  publisher = {American Physical Society (APS)},
}

@Article{Spiess2023,
  author   = {Spiess, Christopher and Töpfer, Sebastian and Sharma, Sakshi and Kržič, Andrej and Cabrejo-Ponce, Meritxell and Chandrashekara, Uday and Döll, Nico Lennart and Rieländer, Daniel and Steinlechner, Fabian},
  journal  = {Physical Review Applied},
  title    = {Clock {Synchronization} with {Correlated} {Photons}},
  year     = {2023},
  issn     = {2331-7019},
  month    = may,
  number   = {5},
  pages    = {054082},
  volume   = {19},
  doi      = {10.1103/PhysRevApplied.19.054082},
  language = {en},
  url      = {https://link.aps.org/doi/10.1103/PhysRevApplied.19.054082},
  urldate  = {2025-03-17},
}

@Article{Komar2014,
  author   = {Kómár, Peter and Kessler, Eric M. and Bishof, Michael and Jiang, Liang and Sørensen, Anders S. and Ye, Jun and Lukin, Mikhail D.},
  journal  = {Nature Physics},
  title    = {A quantum network of clocks},
  year     = {2014},
  issn     = {1745-2473, 1745-2481},
  month    = aug,
  note     = {arXiv:1310.6045 [quant-ph]},
  number   = {8},
  pages    = {582--587},
  volume   = {10},
  doi      = {10.1038/nphys3000},
  url      = {http://arxiv.org/abs/1310.6045},
  urldate  = {2025-04-08},
}

@Misc{Spiess2023a,
  author    = {Spiess, Christopher and Steinlechner, Fabian},
  month     = oct,
  note      = {arXiv:2212.12589 [quant-ph]},
  title     = {Clock synchronization with pulsed single photon sources},
  year      = {2023},
  doi       = {10.48550/arXiv.2212.12589},
  issn      = {2058-9565},
  journal   = {Quantum Science and Technology},
  number    = {1},
  pages     = {015019},
  publisher = {arXiv},
  url       = {http://arxiv.org/abs/2212.12589},
  urldate   = {2025-04-08},
  volume    = {9},
}

@Article{Jozsa2000,
  author    = {Jozsa, Richard and Abrams, Daniel S. and Dowling, Jonathan P. and Williams, Colin P.},
  journal   = {Physical Review Letters},
  title     = {Quantum {Clock} {Synchronization} {Based} on {Shared} {Prior} {Entanglement}},
  year      = {2000},
  issn      = {0031-9007, 1079-7114},
  month     = aug,
  number    = {9},
  pages     = {2010--2013},
  volume    = {85},
  copyright = {http://link.aps.org/licenses/aps-default-license},
  doi       = {10.1103/PhysRevLett.85.2010},
  language  = {en},
  url       = {https://link.aps.org/doi/10.1103/PhysRevLett.85.2010},
  urldate   = {2025-04-08},
}

@Article{Chuang2000,
  author    = {Chuang, Isaac L.},
  journal   = {Physical Review Letters},
  title     = {Quantum {Algorithm} for {Distributed} {Clock} {Synchronization}},
  year      = {2000},
  issn      = {0031-9007, 1079-7114},
  month     = aug,
  number    = {9},
  pages     = {2006--2009},
  volume    = {85},
  copyright = {http://link.aps.org/licenses/aps-default-license},
  doi       = {10.1103/PhysRevLett.85.2006},
  language  = {en},
  url       = {https://link.aps.org/doi/10.1103/PhysRevLett.85.2006},
  urldate   = {2025-04-08},
}

@Article{Dai2020,
  author    = {Dai, Hui and Shen, Qi and Wang, Chao-Ze and Li, Shuang-Lin and Liu, Wei-Yue and Cai, Wen-Qi and Liao, Sheng-Kai and Ren, Ji-Gang and Yin, Juan and Chen, Yu-Ao and Zhang, Qiang and Xu, Feihu and Peng, Cheng-Zhi and Pan, Jian-Wei},
  journal   = {Nature Physics},
  title     = {Towards satellite-based quantum-secure time transfer},
  year      = {2020},
  issn      = {1745-2481},
  month     = may,
  note      = {MAG ID: 3099335333},
  number    = {8},
  pages     = {848--852},
  volume    = {16},
  doi       = {10.1038/s41567-020-0892-y},
  publisher = {Springer Science and Business Media LLC},
}

@Article{Quan2016,
  author    = {Runai Quan and Yiwei Zhai and Mengmeng Wang and Feiyan Hou and Shaofeng Wang and Xiao Xiang and Tao Liu and Shougang Zhang and Ruifang Dong},
  journal   = {Scientific Reports},
  title     = {Demonstration of quantum synchronization based on second-order quantum coherence of entangled photons},
  year      = {2016},
  issn      = {2045-2322},
  note      = {ARXIV\_ID: 1602.06371 MAG ID: 3104084850 S2ID: 68cf0b85aa790186b2a63c8635f3b2666c63c40f},
  number    = {1},
  volume    = {6},
  doi       = {10.1038/srep30453},
  pmcid     = {4958996},
  pmid      = {27452276},
  publisher = {Springer Science and Business Media LLC},
}

@Article{Quan2019,
  author    = {Runai Quan and Ruifang Dong and Yiwei Zhai and Feiyan Hou and Xiao Xiang and Hui Zhou and Chaolin Lv and Zhen Wang and Lixing You and Tao Liu and Shougang Zhang},
  journal   = {Optics Letters},
  title     = {Simulation and realization of a second-order quantum-interference-based quantum clock synchronization at the femtosecond level.},
  year      = {2019},
  issn      = {0146-9592},
  month     = feb,
  note      = {MAG ID: 2914983978 S2ID: a3fa9a6f14675c115bdf8395a21fb3f8db2335e9},
  number    = {3},
  pages     = {614},
  volume    = {44},
  doi       = {10.1364/ol.44.000614},
  pmid      = {30702692},
  publisher = {Optica Publishing Group},
}

@Article{DAuria2020,
  author    = {Virginia D’Auria and Bruno Fedrici and Lutfi Arif Ngah and Florian Kaiser and Laurent Labonté and Olivier Alibart and Sébastien Tanzilli},
  journal   = {npj Quantum Information},
  title     = {A universal, plug-and-play synchronisation scheme for practical quantum networks},
  year      = {2020},
  issn      = {2056-6387},
  month     = jul,
  note      = {MAG ID: 3005385203 S2ID: bf8765168cc5a8ac3861b239df7378bd20eff2aa},
  number    = {1},
  volume    = {6},
  doi       = {10.1038/s41534-020-0245-9},
  publisher = {Springer Science and Business Media LLC},
}

@Article{Ren2012,
  author    = {Ren, Changliang and Hofmann, Holger F.},
  journal   = {Physical Review A},
  title     = {Clock synchronization using maximal multipartite entanglement},
  year      = {2012},
  issn      = {1094-1622},
  month     = jul,
  note      = {MAG ID: 2026278662},
  number    = {1},
  pages     = {014301},
  volume    = {86},
  doi       = {10.1103/physreva.86.014301},
  publisher = {American Physical Society (APS)},
}

@Article{Ho2009,
  author    = {Ho, Caleb and Lamas-Linares, Antia and Kurtsiefer, Christian},
  journal   = {New Journal of Physics},
  title     = {Clock synchronization by remote detection of correlated photon pairs},
  year      = {2009},
  issn      = {1367-2630},
  month     = apr,
  note      = {ARXIV\_ID: 0901.3203 MAG ID: 2042135696 S2ID: 0391b3ead04255ff98711686994986ed562d486b},
  number    = {4},
  pages     = {045011},
  volume    = {11},
  doi       = {10.1088/1367-2630/11/4/045011},
  publisher = {IOP Publishing},
}

@Article{Giovannetti2004,
  author    = {Giovannetti, Vittorio and Lloyd, Seth and Maccone, Lorenzo and Shapiro, Jeffrey H. and Wong, Franco N. C.},
  journal   = {Physical Review A},
  title     = {Conveyor-belt clock synchronization},
  year      = {2004},
  issn      = {1094-1622},
  month     = oct,
  note      = {MAG ID: 2073305537},
  number    = {4},
  pages     = {043808},
  volume    = {70},
  doi       = {10.1103/physreva.70.043808},
  publisher = {American Physical Society (APS)},
}

@Article{Nande2025,
  author    = {Nande, Swaraj Shekhar and Habibie, Muhammad Idham and Ghadimi, Milad and Garbugli, Andrea and Kondepu, Koteswararao and Bassoli, Riccardo and Fitzek, Frank H. P.},
  journal   = {Scientific Reports},
  title     = {Integrating quantum synchronization in future generation networks},
  year      = {2025},
  issn      = {2045-2322},
  month     = mar,
  note      = {Publisher: Nature Publishing Group},
  number    = {1},
  pages     = {7617},
  volume    = {15},
  copyright = {2025 The Author(s)},
  doi       = {10.1038/s41598-025-92038-0},
  language  = {en},
  publisher = {Springer Science and Business Media LLC},
  url       = {https://www.nature.com/articles/s41598-025-92038-0},
  urldate   = {2025-05-06},
}

@InProceedings{Han2019,
  author    = {Han, Mingyu and Crossley, Peter},
  title     = {Vulnerability of IEEE 1588 under Time Synchronization Attacks},
  year      = {2019},
  month     = aug,
  note      = {ISSN: 1944-9933},
  pages     = {1--5},
  publisher = {IEEE},
  doi       = {10.1109/PESGM40551.2019.8973494},
  journal   = {2019 IEEE Power &amp; Energy Society General Meeting (PESGM)},
  url       = {https://ieeexplore.ieee.org/document/8973494/},
  urldate   = {2025-08-11},
}

@Article{Itkin2020,
  author    = {Itkin, Eyal and Wool, Avishai},
  journal   = {IEEE Transactions on Dependable and Secure Computing},
  title     = {A Security Analysis and Revised Security Extension for the Precision Time Protocol},
  year      = {2020},
  issn      = {2160-9209},
  month     = jan,
  number    = {1},
  pages     = {22--34},
  volume    = {17},
  doi       = {10.1109/TDSC.2017.2748583},
  publisher = {Institute of Electrical and Electronics Engineers (IEEE)},
  url       = {https://ieeexplore.ieee.org/document/8025399/},
  urldate   = {2025-08-27},
}

@Article{Lee2019a,
  author    = {Lee, Jianwei and Shen, Lijiong and Cerè, Alessandro and Troupe, James and Lamas-Linares, Antia and Kurtsiefer, Christian},
  journal   = {Applied Physics Letters},
  title     = {Asymmetric delay attack on an entanglement-based bidirectional clock synchronization protocol},
  year      = {2019},
  issn      = {1077-3118},
  month     = sep,
  note      = {MAG ID: 2978911397},
  number    = {14},
  volume    = {115},
  doi       = {10.1063/1.5121489},
  publisher = {AIP Publishing},
}

@Misc{AndreoliFang2018,
  author   = {Andreoli-Fang, Jennifer},
  month    = jul,
  title    = {Introducing the {DOCSIS} {Synchronization} {Techniques} {Specification}},
  year     = {2018},
  journal  = {CableLabs},
  language = {en},
  url      = {https://www.cablelabs.com/blog/introducing-docsis-synchronization-techniques-specification},
  urldate  = {2025-09-02},
}

@Misc{Kong2017,
  author    = {Kong, Xiangyu and Xin, Tao and Wei, ShiJie and Wang, Bixue and Wang, Yunzhao and Li, Keren and Long, GuiLu},
  month     = aug,
  note      = {arXiv:1708.06050 [quant-ph]},
  title     = {Demonstration of multiparty quantum clock synchronization},
  year      = {2017},
  doi       = {10.48550/arXiv.1708.06050},
  issn      = {1570-0755},
  journal   = {Quantum Information Processing},
  number    = {11},
  publisher = {Springer Science and Business Media LLC},
  url       = {http://arxiv.org/abs/1708.06050},
  urldate   = {2025-09-12},
  volume    = {17},
}

@Misc{Hong2021,
  author    = {Hong, Huibo and Quan, Runai and Xiang, Xiao and Xue, Wenxiang and Quan, Honglei and Zhao, Wenyu and Liu, Yuting and Cao, Mingtao and Liu, Tao and Zhang, Shougang and Dong, Ruifang},
  month     = oct,
  note      = {arXiv:2111.00380 [quant-ph]},
  title     = {Demonstration of 50 km Fiber-Optic Two-Way Quantum Clock Synchronization},
  year      = {2021},
  doi       = {10.48550/arXiv.2111.00380},
  issn      = {0733-8724},
  journal   = {Journal of Lightwave Technology},
  number    = {12},
  pages     = {3723-3728},
  publisher = {arXiv},
  url       = {http://arxiv.org/abs/2111.00380},
  urldate   = {2025-09-15},
  volume    = {40},
}

@Article{Giovannetti2001,
  author    = {Giovannetti, Vittorio and Lloyd, Seth and Maccone, Lorenzo},
  journal   = {Nature},
  title     = {Quantum-enhanced positioning and clock synchronization},
  year      = {2001},
  issn      = {0028-0836, 1476-4687},
  month     = jul,
  number    = {6845},
  pages     = {417--419},
  volume    = {412},
  copyright = {http://www.springer.com/tdm},
  doi       = {10.1038/35086525},
  language  = {en},
  url       = {https://www.nature.com/articles/35086525},
  urldate   = {2025-09-29},
}

@Misc{Crum2025,
  author    = {Crum, Noah and Hassan, Md Mehdi and Siopsis, George},
  month     = sep,
  note      = {arXiv:2510.00199 [quant-ph]},
  title     = {Practical Quantum Clock Synchronization Using Weak Coherent Pulses},
  year      = {2025},
  copyright = {Creative Commons Attribution Non Commercial Share Alike 4.0 International},
  doi       = {10.48550/arXiv.2510.00199},
  publisher = {arXiv},
  url       = {http://arxiv.org/abs/2510.00199},
  urldate   = {2025-10-03},
}

@Article{Alghamdi2022,
  author   = {Alghamdi, Waleed and Schukat, Michael},
  journal  = {Sensors (Basel, Switzerland)},
  title    = {A {Security} {Enhancement} of the {Precision} {Time} {Protocol} {Using} a {Trusted} {Supervisor} {Node}},
  year     = {2022},
  issn     = {1424-8220},
  month    = may,
  note     = {tex.pmcid: PMC9147087},
  number   = {10},
  pages    = {3671},
  volume   = {22},
  doi      = {10.3390/s22103671},
  pmid     = {35632078},
  url      = {https://www.ncbi.nlm.nih.gov/pmc/articles/PMC9147087/},
  urldate  = {2025-08-26},
}

@InProceedings{Ullmann2009,
  author    = {Ullmann, Markus and Vögeler, Matthias},
  booktitle = {Control and {Communication} 2009 {International} {Symposium} on {Precision} {Clock} {Synchronization} for {Measurement}},
  title     = {Delay attacks — Implication on NTP and PTP time synchronization},
  year      = {2009},
  month     = oct,
  note      = {ISSN: 1949-0313},
  pages     = {1--6},
  publisher = {IEEE},
  doi       = {10.1109/ISPCS.2009.5340224},
  journal   = {2009 International Symposium on Precision Clock Synchronization for Measurement, Control and Communication},
  url       = {https://ieeexplore.ieee.org/document/5340224},
  urldate   = {2025-10-24},
}

@article{Sandholm2026,
  title={Lightspeed Data Compute for the Space Era},
  author={Sandholm, Thomas and Huberman, Bernardo A and Segeljakt, Klas and Carbone, Paris},
  journal={arXiv preprint arXiv:2601.17589},
  year={2026}
}

@article{Kwiat1995,
  title={New high-intensity source of polarization-entangled photon pairs},
  author={Kwiat, Paul G and Mattle, Klaus and Weinfurter, Harald and Zeilinger, Anton and Sergienko, Alexander V and Shih, Yanhua},
  journal={Physical Review Letters},
  volume={75},
  number={24},
  pages={4337},
  year={1995},
  publisher={APS}
}

@Article{Glauber1963,
  author    = {Glauber, Roy J.},
  journal   = {Physical Review},
  title     = {The Quantum Theory of Optical Coherence},
  year      = {1963},
  issn      = {0031-899X},
  month     = jun,
  number    = {6},
  pages     = {2529--2539},
  volume    = {130},
  copyright = {http://link.aps.org/licenses/aps-default-license},
  doi       = {10.1103/PhysRev.130.2529},
  language  = {en},
  publisher = {American Physical Society (APS)},
  url       = {https://link.aps.org/doi/10.1103/PhysRev.130.2529},
  urldate   = {2025-11-13},
}

@Article{Burnham1970,
  author    = {Burnham, David C. and Weinberg, Donald L.},
  journal   = {Physical Review Letters},
  title     = {Observation of {Simultaneity} in {Parametric} {Production} of {Optical} {Photon} {Pairs}},
  year      = {1970},
  issn      = {0031-9007},
  month     = jul,
  number    = {2},
  pages     = {84--87},
  volume    = {25},
  copyright = {http://link.aps.org/licenses/aps-default-license},
  doi       = {10.1103/PhysRevLett.25.84},
  language  = {en},
  url       = {https://link.aps.org/doi/10.1103/PhysRevLett.25.84},
  urldate   = {2025-11-19},
}

@Article{Mills1991,
  author     = {Mills, D.L.},
  journal    = {IEEE Transactions on Communications},
  title      = {Internet time synchronization: the network time protocol},
  year       = {1991},
  issn       = {1558-0857},
  month      = oct,
  number     = {10},
  pages      = {1482--1493},
  volume     = {39},
  doi        = {10.1109/26.103043},
  shorttitle = {Internet time synchronization},
  url        = {https://ieeexplore.ieee.org/document/103043/},
  urldate    = {2025-11-19},
}

@Article{Lewandowski1999,
  author     = {Lewandowski, W. and Azoubib, J. and Klepczynski, W.J.},
  journal    = {Proceedings of the IEEE},
  title      = {{GPS}: primary tool for time transfer},
  year       = {1999},
  issn       = {1558-2256},
  month      = jan,
  number     = {1},
  pages      = {163--172},
  volume     = {87},
  doi        = {10.1109/5.736348},
  shorttitle = {{GPS}},
  url        = {https://ieeexplore.ieee.org/document/736348/},
  urldate    = {2025-11-19},
}

@Article{Hong2024,
  author    = {Hong, Huibo and Quan, Runai and Xiang, Xiao and Liu, Yuting and Liu, Tao and Cao, Mingtao and Dong, Ruifang and Zhang, Shougang},
  journal   = {Journal of Lightwave Technology},
  title     = {Quantum Two-Way Time Transfer Over a 103 km Urban Fiber},
  year      = {2024},
  issn      = {1558-2213},
  month     = mar,
  number    = {5},
  pages     = {1479--1486},
  volume    = {42},
  doi       = {10.1109/JLT.2023.3323434},
  publisher = {Institute of Electrical and Electronics Engineers (IEEE)},
  url       = {https://ieeexplore.ieee.org/document/10274994/},
  urldate   = {2025-11-19},
}

@InProceedings{Lipinski2011,
  author     = {Lipinski, Maciej and Wlostowski, Tomasz and Serrano, Javier and Alvarez, Pablo},
  booktitle  = {2011 IEEE International Symposium on Precision Clock Synchronization for Measurement, Control and Communication},
  title      = {White rabbit: a PTP application for robust sub-nanosecond synchronization},
  year       = {2011},
  address    = {Munich, Germany},
  month      = sep,
  pages      = {25--30},
  publisher  = {IEEE},
  doi        = {10.1109/ISPCS.2011.6070148},
  isbn       = {978-1-61284-893-8},
  language   = {en},
  shorttitle = {White rabbit},
  url        = {http://ieeexplore.ieee.org/document/6070148/},
  urldate    = {2025-11-20},
}

@Article{Ducoing2025,
  author    = {Ducoing, Sage and Agullo, Ivan and Troupe, James E and Haldar, Stav},
  journal   = {Physical Review Applied},
  title     = {Quantum-assisted master clock in the sky: Global synchronization from satellites at subnanosecond precision},
  year      = {2025},
  number    = {1},
  pages     = {014052},
  volume    = {23},
  publisher = {APS},
}

@Book{Boyd2020,
  author    = {Boyd, Robert W.},
  publisher = {Elsevier},
  title     = {Nonlinear Optics},
  year      = {2020},
  edition   = {Fourth},
  isbn      = {9780128110027},
  month     = mar,
  doi       = {10.1016/c2015-0-05510-1},
}
\bibliographystyle{unsrt}

\end{document}